\documentclass{iopart}

\usepackage{iopams}
\usepackage{simplewick}
\usepackage{graphicx}

\usepackage[cp1251]{inputenc}
\usepackage[T2A]{fontenc}
\usepackage[russian,english]{babel}

\usepackage[english]{babel}
\usepackage[sort]{cite}

\newcommand{\Sp}{\mathop{\mathrm{Sp}}}
\renewcommand{\Re}{\mathop{\mathrm{Re}}}
\renewcommand{\Im}{\mathop{\mathrm{Im}}}

\begin{document}
\title[Spin-fluctuation theory beyond Gaussian approximation]{Spin-fluctuation theory
beyond Gaussian approximation}

\author{N B Melnikov$^1$\footnote{Also at
the Central Economics and Mathematics Institute of the Russian Academy of Sciences,
117418 Moscow, Russia}, B I Reser$^2$ and V I Grebennikov$^2$}
\address{$^1$ Moscow State University, 119992 Moscow, Russia}
\address{$^2$ Institute of Metal Physics, Ural Branch of the Russian \\
\hspace{0.15cm} Academy of Sciences, 620041 Ekaterinburg, Russia}
\eads{\mailto{melnikov@cs.msu.su}, \mailto{reser@imp.uran.ru},
\mailto{greben@imp.uran.ru}}

\begin{abstract}
A characteristic feature of the Gaussian approximation in the functional-integral
approach to the spin-fluctuation theory is the jump phase transition to the
paramagnetic state. We eliminate the jump and obtain a continuous second-order
phase transition by taking into account high-order terms in the expansion of the
free energy in powers of the fluctuating exchange field. The third-order term of
the free energy renormalizes the mean field, and fourth-order term, responsible
for the interaction of the fluctuations, renormalizes the spin susceptibility. The
extended theory is applied to the calculation of magnetic properties of Fe-Ni Invar.
\end{abstract}


\vspace{1cm}

This is an author-created, un-copyedited version of an article accepted for publication in \emph{Journal of Physics A: Mathematical and Theoretical}. IOP Publishing Ltd is not responsible for any errors or omissions in this version of the manuscript or any version derived from it. The definitive publisher authenticated version is available online at http://stacks.iop.org/1751-8121/43/195004.

\maketitle

\section{Introduction}
Fluctuations of the electron spin density play a predominant role in the
thermodynamics of ferromagnetic metals (see, e.g., \cite{Mor85} and references
therein). Most of the progress in the spin-fluctuation theory (SFT) has been
achieved within the functional-integral approach \cite{Str57,Hub59}. Using
the single-site and static approximation, Hubbard~\cite{Hub79},
Hasegawa~\cite{Has79} and Grebennikov \emph{et al.}~\cite{GPS81} obtained
a quantitative description of magnetic properties at finite temperatures,
which was much better than in the Stoner mean-field theory (\fref{spin-configs},
left). To go beyond the single-site approximation Hertz and Klenin~\cite{HK74}
suggested a self-consistent Gaussian approximation. However, it used the static
long-wave limit and was restricted to paramagnets. To ferromagnetic metals,
the variational approach~\cite{HK74} was extended by Grebennikov~\cite{Gre88}
who took into account dynamics and space correlations of the fluctuations,
the latter again only in the paramagnetic state. A complete dynamic non-local
approximation to the SFT for ferromagnetic metals was developed by Reser and
Grebennikov~\cite{RG97}.

The Gaussian SFT~\cite{RG97} gives a good agreement with experiment over
a wide range of temperatures. However, at high temperatures the Gaussian
approximation yields a discontinuous change in magnetic characteristics
(see~\cite{RM08} and references therein). The main reason for the first-order
phase transition is that Gaussian approximation implies independent
`harmonic' fluctuations of the spin density and thus fails to account
for their interaction.

The first-order phase transition has been observed in various versions
of the self-consistent renormalization (SCR) theory of spin fluctuations,
developed for weak ferromagnetic metals (for a review, see \cite{Mor85}).
Particularly, in~\cite{Lon85} it was argued that the first-order discontinuity
in the SCR theory can be eliminated by taking into account the rotational
invariance of the system. This leads to two integro-differential equations for
the longitudinal $\chi_\parallel$ and transverse $\chi_\perp$ susceptibilities.
A simple relation that couples $\chi_\parallel$ and $\chi_\perp$ was suggested
in~\cite{Tak86} from the assumption that the total local spin fluctuation,
i.e. the sum of the zero-point and thermal spin fluctuations, is conserved,
as it is in the Heisenberg local moment theory and may be somehow justified
for weak ferromagnets.

In the present paper, we improve the coupling of the fluctuations in
the Gaussian SFT~\cite{RG97} by taking into account high-order terms of
the free energy of electrons $F(V)$ in the fluctuating exchange field~$V$
(see~\cite{RMG09} for a brief summary). First, in the fourth-order Taylor
expansion of the free energy $F(V)$, we take a partial average with respect
to $\Delta V$ in the third- and fourth-order terms replacing them by a linear
and quadratic terms, respectively. Adding these corrections to the Taylor
terms of the first- and second-order, we come to the \emph{extended} function
$F(V)$. The best quadratic approximation is constructed as
in~\cite{HK74,Gre88,RG97}, with the help of the free energy minimum
principle~\cite{Fey55}, but using the first- and second-order derivatives of
the \emph{extended} function $F(V)$. In the final computational formulae,
the third-order term renormalizes the mean field, and fourth-order term
renormalizes the susceptibility (this includes the Gaussian SFT~\cite{RG97}
as a special case with the renormalizations set to unity).

The fundamental difference between our treatment of the high-order terms
and the previous ones is that in our approach the ferromagnetic state is
changed self-consistently (for treatments of the fourth-order term in
the paramagnetic state see \cite{Mor85,SEW71,Her76,Sta85} and references
therein). Another advantage of our approximation to the SFT is that the
integral equation for the mean Green function (coherent potential equation)
is finally reduced to a system of four nonlinear equations with four unknowns,
which is only slightly more complex than the Stoner mean-field theory. 
It is significant to note that, solving the coherent potential equation 
directly requires a number of additional simplifications, such as neglect of 
rotational invariance and the mode-mode `frequency' interactions, and, most 
important, the single-site approximation \cite{Kak02,KF08}.

The extended Gaussian approximation of the SFT is applied to numerical
calculations of magnetic properties of $\mathrm{Fe}_{0.65}\mathrm{Ni}_{0.35}$
Invar alloy at finite temperatures. This alloy has been intensively studied
recently (see, e.g., \cite{RKM07} and references therein), but mostly at zero
temperature, i.e. without the quantum statistics. Our choice of the Fe-Ni
Invar to illustrate the possibilities of the extended SFT is motivated by
problems of temperature dependence which were found in the quantum-statistical
treatment of this Invar \cite{Res04a, Res04b, Res07, RM08}. (Obviously, before
that the new method has been tested on a simplified clean system,
such as elemental Fe.)

It is known that the Fe-Ni Invar is a complex disordered system. However,
the comparison of the calculation results for the disordered alloy
$\mathrm{Fe}_{0.65}\mathrm{Ni}_{0.35}$ \cite{Res07, RM08} and ordered
compound $\mathrm{Fe}_3\mathrm{Ni}$ \cite{Res04a, Res04b} showed that
the effect of disorder in the filling of sites with Fe and Ni atoms on
the magnetic properties of the Fe-Ni Invar is insignificant. This
conclusion agrees with earlier results for the ordered and disordered
$\mathrm{Fe}_{0.72}\mathrm{Pt}_{0.28}$ Invars
(see, e.g.,~\cite{Shi94}, table~10--1). The weak influence of the atomic
disorder on the magnetic properties of the Fe-Ni Invar at finite temperatures
is explained by the \emph{integral} dependence on the electronic energy
structure in all the equations of the SFT. The details of the initial
density of states (DOS) do not exert the decisive effect on the results
of the calculation.

\section{Quadratic approximation taking into account high-order terms}
\subsection{Free energy of electrons in a fluctuating field}
\begin{figure}[hbt]
\begin{center}
\includegraphics[width=0.75\textwidth]{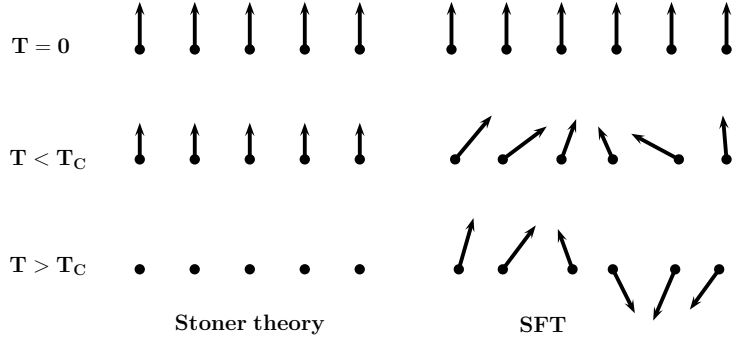}
\end{center}
\caption{\label{spin-configs} A schematic picture of exchange
field configurations in \\ the Stoner mean-field theory and SFT.}
\end{figure}

The Stratonovich-Hubbard transformation \cite{Str57,Hub59} replaces the
pair interaction of electrons by the interaction of the electrons
with the exchange field\footnote{We neglect the charge field, same
as in \cite{RG97}.}
\begin{equation}\label{V-def}
V \equiv (V_1, V_2, ...),
\qquad
V_j = \sum_\alpha V_j^\alpha (\tau)\sigma^\alpha,
\end{equation}
fluctuating in space (see~\fref{spin-configs}, right)
and in
`time' $\tau \in [0, 1/T]$ ($j$ being the site number,
$\sigma^\alpha$ the Pauli matrix, $\alpha = x, y, z$, and $T$ the
temperature, in energy units).
Hence the partition function can be written as a functional integral
\begin{equation}\label{Z}
\fl
 Z = \rme^{-\mathcal{F}/T}
 = \biggl( \int \rme^{-F_0(V)/T}\, DV(\tau) \biggr)^{-1}
 \int \rme^{-F_0(V)/T}\, \rme^{-F_1(V)/T}\, DV(\tau),
\end{equation}
where
\begin{equation}\label{F0-def}
    F_0(V) = N_\rmd T \int_0^{1/T} \sum_{j}
    \Sp \Biggl( \frac{V_{j}^2(\tau)}{2U} \Biggr) \rmd\tau
    \equiv \Tr \biggl( \frac{V^2}{2U} \biggr)
\end{equation}
is the energy of the exchange field, and
\begin{equation}\label{F1-def}
    F_1(V) = T \Tr \ln G(V)
\end{equation}
is the free energy of non-interacting electrons in the
field $V$ expressed in terms of the Green function
\begin{equation}\label{G-def}
  G(V) = (z + \mu - H_0 - V)^{-1}.
\end{equation}
Here $N_\rmd=5$ is the number of \emph{degenerate} d bands, $U$ is the
single-site electron-electron interaction constant, $z$ is the
energy variable, $\mu$ is the chemical potential, and $H_0$ is the
sum of kinetic and potential energy of the non-interacting band electrons.
Symbol $\Sp$ denotes
the trace over spin indices, and $\Tr$ stands for the full matrix
trace, independent of the particular representation. To simplify
the notation, we omit the band index and write the prefactor $N_\rmd$
in the trace $\Tr$. In expression \eref{F1-def}, we omit the terms
independent of the field $V$ (for details, see~\ref{Appendix-A}).

\subsection{Partial averaging of high order terms of the free energy}
Since $H_0$ and $V$ cannot be diagonalized simultaneously in either
coordinate-`time' or momentum-`frequency' spaces, exact
formulae \eref{F1-def} and \eref{G-def} are of little use to
calculate the partition function \eref{Z} without an appropriate
approximation. Therefore, integral \eref{Z} is replaced by
a Gaussian integral or, equivalently, the exact expression
\begin{equation}\label{F}
    F(V) = F_0(V) + F_1(V)
    = T \Tr \biggl( \frac{V^2}{2UT} + \ln G(V) \biggr)
\end{equation}
is replaced by a \emph{quadratic} form:
\begin{equation}\label{F-to-F2}
   F(V) \to  F^{(2)}(V) = \Tr \bigl( \Delta V A \Delta V \bigr),
   \qquad
   \Delta V = V - \bar V.
\end{equation}
Hence, we arrive to the Gaussian fluctuating field $V$ with the
probability density
\begin{equation}\label{matrix-Gauss-density}
  p(V) \propto \exp \biggl( -\frac1T
  \Tr \bigl( \Delta V A \Delta V \bigr) \biggr),
\end{equation}
the mean field $\bar V$ and matrix $A$ being the unknown
parameters of the SFT to be determined. Compared to conventional
Gaussian SFTs, we take into account the `anharmonicity' of the
fluctuations by renormalizing the parameters $\bar V$ and $A$.

First, we expand function \eref{F} in Taylor series to the fourth
order in $\Delta V = V - \tilde V$:
\begin{eqnarray}\label{F-4th-order}
 \fl F(V) \approx F(\tilde V)
    + T \Tr \biggl( \frac{\tilde V \Delta V}{UT}
                  + G(\tilde V) \Delta V \biggr)
    + \frac12 T \Tr \biggl( \frac{\Delta V^2}{UT}
    + \bigl(G(\tilde V) \Delta V \bigr)^2 \biggr) \nonumber \\
    + \frac13 T \Tr \bigl(G(\tilde V) \Delta V\bigr)^3
    + \frac14 T \Tr \bigl(G(\tilde V) \Delta V\bigr)^4,
\end{eqnarray}
where $\tilde V$ is some \emph{arbitrary} value of the exchange
field. For the third-order term, we define the Gaussian partial
averaging by the formula
\begin{eqnarray}\label{V-decoupling3}
\fl
    \Tr \bigl( G\Delta V G\Delta V G\Delta V\bigr) \approx
    \Tr \bigl(\bcontraction{G}{\Delta V}{G}{\Delta V}
        G\Delta V G\Delta V G\Delta V\bigr)
    + \Tr \bigl(\bcontraction{G}{\Delta V}{G\Delta V G}{\Delta V}
          G\Delta V G\Delta V G\Delta V\bigr) \nonumber \\
    + \Tr \bigl(\bcontraction{G\Delta V G}{\Delta V}{G}{\Delta V}
          G\Delta V G\Delta V G\Delta V\bigr)
    = 3 \Tr \bigl(\bcontraction{G}{\Delta V}{G}{\Delta V}
          G\Delta V G\Delta V G\Delta V\bigr)
\end{eqnarray}
and for the fourth-order term by the formula
\begin{eqnarray}\label{V-decoupling4-raw}
\fl
    \Tr \bigl(G\Delta V G\Delta V G\Delta V G\Delta V\bigr) \nonumber \\
    \approx 4\Tr \bigl(\bcontraction{G}{\Delta V}{G}{\Delta V}
                       G\Delta V G\Delta V G\Delta V G\Delta V\bigr)
    + 2\Tr \bigl(\bcontraction{G}{\Delta V}{G\Delta V G}{\Delta V}
                 G\Delta V G\Delta V G\Delta V G\Delta V\bigr) \nonumber \\
    - 2\Tr \bigl(\bcontraction{G}{\Delta V}{G}{\Delta V}
      \bcontraction{G\Delta V G\Delta V G}{\Delta V}{G}{\Delta V}
                    G\Delta V G\Delta V G\Delta V G\Delta V\bigr)
    - \Tr \bigl(\bcontraction{G}{\Delta V}{G\Delta V G}{\Delta V}
      \bcontraction[2ex]{G\Delta V G}{\Delta V}{G\Delta V G}{\Delta V}
                         G\Delta V G\Delta V G\Delta V G\Delta V\bigr),
\end{eqnarray}
where the underbrace denotes the averaging with
the Gaussian probability density~\eref{matrix-Gauss-density}:
\begin{equation*}\label{V-average}
  \bigl(\dots \bcontraction{}{\Delta V}{\dots}{\Delta V}
        \Delta V \dots \Delta V \dots\bigr)
  = \int \bigl(\dots \Delta V \dots \Delta V \dots \bigr) p(V)\, DV.
\end{equation*}

In formulae \eref{V-decoupling3} and \eref{V-decoupling4-raw}, the
combinatorial multipliers correspond to the number of all
possible pairings of the terms $\Delta V$ with the cyclic invariance of
the trace taken into account. In formula \eref{V-decoupling4-raw},
the last two terms are chosen so that both sides have equal mean values.
These terms will be omitted for brevity since they do not contain
the variable $\Delta V$ and lead only to a change of unimportant
free term in expansion \eref{F-4th-order}. Furthermore, we approximate
the second term in the right-hand side of~\eref{V-decoupling4-raw}
by the first one. Thus, relation \eref{V-decoupling4-raw} takes the form
\begin{eqnarray}\label{V-decoupling4}
    \Tr \bigl(G\Delta V G\Delta V G\Delta V G\Delta V\bigr)
    \approx 6\Tr \bigl(\bcontraction{G}{\Delta V}{G}{\Delta V}
                 G\Delta V G\Delta V G\Delta V G\Delta V\bigr).
\end{eqnarray}
Using formulae \eref{V-decoupling3} and \eref{V-decoupling4}, we
transform the third- and fourth-order terms in expansion
\eref{F-4th-order} into corrections to the first- and second-order
terms, respectively:
\begin{eqnarray}\label{F-4th-order-decoupled}
\fl
    F(V) \approx F(\tilde V)
    + T \Tr \biggl( \frac{\tilde V \Delta V}{UT}
    + G(\tilde V)\Delta V
    + \bcontraction{G(\tilde V)}{\Delta V}{G(\tilde V)}{\Delta V}
    G(\tilde V)\Delta V G(\tilde V)\Delta V G(\tilde V)\Delta V \biggr) \\
\fl \qquad
    + \frac12 T \Tr \biggl( \frac{\Delta V^2}{UT}
    + G(\tilde V)\Delta V G(\tilde V)\Delta V
    + 3 \bcontraction{G(\tilde V)}{\Delta V}{G(\tilde V)}{\Delta V}
    G(\tilde V)\Delta V G(\tilde V)\Delta V
    G(\tilde V)\Delta V G(\tilde V)\Delta V \biggr).
\nonumber
\end{eqnarray}

\subsection{Quadratic approximation based on the free energy minimum principle}
Following \cite{HK74,Gre88,RG97}, the best quadratic approximation $F^{(2)}(V)$
is constructed using the free energy minimum principle \cite{Fey55}.

Define the average with the Gaussian density \eref{matrix-Gauss-density}
by the formula
\begin{equation*}
\fl
    \langle \dots \rangle
    = \biggl( \int \exp\bigl( - F^{(2)}(V)/T \bigr)\, DV \biggr)^{-1}
    \int (\dots) \exp\bigl( - F^{(2)}(V)/T \bigr)\, DV,
\end{equation*}
where $F^{(2)}(V)$ is a quadratic function of the form \eref{F-to-F2}.
Then the identity
\begin{equation*}
\fl
    \int \exp\bigl( - F(V)/T \bigr)\, DV
  = \int \exp\Bigl( - \bigl(F(V)-F^{(2)}(V) \bigr)\bigl.\bigr/T \Bigr)
    \exp\bigl( - F^{(2)}(V)/T \bigr)\, DV
\end{equation*}
can be rewritten as
\begin{eqnarray*}
\fl
    \biggl( \int \exp\bigl( - F^{(2)}(V)/T \bigr)\, DV \biggr)^{-1}
    &\int \exp\bigl( - F(V)/T \bigr)\, DV \\
    &= \Bigl\langle \exp\Bigl( - \bigl( F(V)-F^{(2)}(V) \bigr)
       \bigl.\bigr/T\Bigr) \Bigr\rangle.
\end{eqnarray*}
Applying inequality $\langle \exp f \rangle \ge \exp \langle f \rangle$,
$f$ being an arbitrary set of real quantities, and taking the logarithm,
we come to the upper bound for the free energy:
\begin{equation}\label{free-energy-principle}
  \mathcal{F} \le \mathcal{F}^{(2)}
    + \bigl\langle F(V) - F^{(2)}(V) \bigr\rangle,
\end{equation}
where
\begin{eqnarray*}
\fl
  \mathcal{F}  &= - T \ln\int \exp\bigl( -F(V)/T \bigr)\, DV, \qquad
  \mathcal{F}^{(2)} &= - T \ln\int \exp\bigl( -F^{(2)}(V)/T \bigr)\, DV .
\end{eqnarray*}

To get the `best' approximation in the class of all quadratic functions
\eref{F-to-F2}, one minimizes the right-hand side of \eref{free-energy-principle}
to obtain the equation (for details, see~\cite{MR10})
\begin{equation}\label{linear-term}
     \left\langle \frac{\partial F(V)}{\partial V} \right\rangle = 0,
\end{equation}
so that $\tilde V$ is equal to the mean field
$\bar V \equiv \langle \tilde V \rangle$.
The matrix of the quadratic form is given by
\begin{equation}\label{quadratic-term}
     A = \frac12 \left\langle \frac{\partial^2 F(V)}{\partial V^2} \right\rangle.
\end{equation}

In the present paper, we apply the formulae \eref{linear-term} and
\eref{quadratic-term} to the modified function \eref{F-4th-order-decoupled}
instead of the original free energy \eref{F}, as it was done in \cite{RG97}.
In expression \eref{F-4th-order-decoupled}, we average over $\tilde V$ with the
Gaussian density \eref{matrix-Gauss-density}, everywhere but in $\Delta V$,
and replace the $\Delta V = V - \tilde V$ terms by $\Delta V = V - \bar V$.
The averaged linear term in \eref{F-4th-order-decoupled} annihilates:
\begin{equation}\label{HK1}
\fl
    T \Tr \biggl(  \frac{\bar V \Delta V }{UT}
    + \bigl\langle G(\tilde V)\bigr\rangle \Delta V
    + \bigl\langle \bcontraction{G(\tilde V)}{\Delta V}{G(\tilde V)}{\Delta V}
    G(\tilde V)\Delta V G(\tilde V)\Delta V G(\tilde V) \bigr\rangle \Delta V \biggr)
    = 0
\end{equation}
identically over $\Delta V$. Hence
the quadratic form $F^{(2)}(V)$ contains only the second-order
term:
\begin{eqnarray}\label{HK2}
\fl
    F(V) \approx F^{(2)}(V)
    = \frac12 T \Tr \biggl( \frac{\Delta V^2}{UT}
    &+ \bigl\langle G(\tilde V)\Delta V G(\tilde V)\Delta V \bigr\rangle \nonumber \\
    &+ 3 \bigl\langle \bcontraction{G(\tilde V)}{\Delta V}{G(\tilde V)}{\Delta V}
    G(\tilde V)\Delta V G(\tilde V)\Delta V
    G(\tilde V)\Delta V G(\tilde V)\Delta V \bigr\rangle \biggr).
\end{eqnarray}

To simplify expressions \eref{HK1} and \eref{HK2} one step further, we introduce
yet another partial averaging:
\begin{eqnarray*}
  G \bcontraction{}{\Delta V}{G}{\Delta V} \Delta V
  G \Delta V G \Delta V &\approx
    \bigl(\Tr(\mathbf{1})\bigr)^{-1}
    \Tr \Bigl( G \bcontraction{}{\Delta V}{G}{\Delta V}
    \Delta V G \Delta V \Bigr)\, G \Delta V \label{Tr3}.
\end{eqnarray*}
Finally, we replace the average over $\tilde V$ of the product of the
Green functions $G(\tilde V)$  by the product of mean Green
functions $\langle G(\tilde V) \rangle \equiv \bar G$.
Hence relation \eref{HK1} reduces to
\begin{equation}\label{HK1-decoupled}
    U^{-1} \Tr \bigl( \bar V \Delta V \bigr)
    + \bigl(1 + \eta \bigr) T \Tr \bigl( \bar G \Delta V \bigr) = 0
\end{equation}
and quadratic form \eref{HK2} transforms to
\begin{equation}\label{HK2-decoupled}
    F^{(2)}(V) = \frac1{2U} \Tr \bigl( \Delta V^2 \bigr)
    + \frac12 \bigl(1 + 3 \eta \bigr)
    T \Tr \bigl( \bar G \Delta V \bar G \Delta V \bigr),
\end{equation}
where the correction coefficient is
\begin{equation}\label{eta-def}
  \eta = \bigl(\Tr(\mathbf{1})\bigr)^{-1}
    \Tr \Bigl(\bar G \bcontraction{}{\Delta V}{\bar G}{\Delta V}
    \Delta V \bar G \Delta V \Bigr).
\end{equation}

\subsection{Equations for the mean field and chemical potential}
In the ferromagnetic state, we choose the $z$-axis along the
direction of the mean field:
\begin{equation*}
    \bar V = \bar V^{z} \sigma^{z}.
\end{equation*}
Then the mean Green function is spin-diagonal due to
\begin{equation}\label{G-xy}
\bar G^x = 0, \qquad \bar G^y = 0.
\end{equation}
Hence, using the well-known relation for the Pauli matrices:
$\Sp(\sigma^{\alpha}\sigma^{\beta})=2\delta_{\alpha\beta}$, we
rewrite equation \eref{HK1-decoupled} as
\begin{equation}\label{mean-field-Tr}
    U^{-1} \Tr ( \bar V^z \Delta V^z) + \bigl(1 + \eta \bigr)
    T \Tr ( \bar G^{z} \Delta V^z) = 0.
\end{equation}

In order to transform \eref{mean-field-Tr} to the mean-field
equation, we use the momentum-`frequency' representation. Since
the exchange field \eref{V-def} is diagonal in the
coordinate-`time' representation, its Fourier transform is
homogeneous:
\begin{equation}\label{V-momentum-frequency}
  V^{\alpha}_{\mathbf{kk}'nn'} =
  V^{\alpha}_{\mathbf{k-k}',\,n-n'},
\end{equation}
where $\mathbf{k}$ is the wave vector, taking values in the
Brillouin zone, and $\omega_n=(2n+1)\pi T$ are the thermodynamic
`frequencies'. Furthermore, the mean field $\bar V$ is a
constant in the coordinate-`time' representation, and hence its
Fourier transform has the single non-zero coefficient:
\begin{equation*}\label{ferro-mean-field}
  \bar V^{\alpha}_{\mathbf{q}m}
  = \bar V^z_{00} \delta_{\mathbf{q}0} \delta_{m0} \delta_{\alpha z}.
\end{equation*}
Thus, the first term in the left-hand side of \eref{mean-field-Tr}
reduces to
\begin{equation}\label{mean-field-eq-Tr1}
    U^{-1}\Tr \bigl(\bar V^{z} \Delta V^{z}\bigr)
    = u^{-1} \bar V^z_{00} \Delta V^{z}_{00},
\end{equation}
where $u=U/N_\rmd$. Similarly, the mean Green function is
transitionally invariant in space and `time': $\bar G_{jj'}
(\tau,\tau') = \bar G_{j-j'}(\tau-\tau')$, hence its Fourier
transform is diagonal:
\begin{equation}\label{G-momentum-frequency}
  \bar G^{\gamma}_{\mathbf{kk}'nn'}
  = \bar G^{\gamma}_{\mathbf{k}n} \delta_{\mathbf{kk}'}\delta_{nn'},
  \qquad
  \gamma = 0,z.
\end{equation}
Therefore, the trace in the second term of \eref{mean-field-Tr} can be written as
\begin{equation}\label{mean-field-eq-Tr2}
     \Tr \bigl( \bar G^{z} \Delta V^{z} \bigr)
    = \sum_{\mathbf{k}n} \bar G^{z}_{\mathbf{k}n} \Delta V^{z}_{00}
    = \Tr (\bar G^{z}) \Delta V^{z}_{00}.
\end{equation}
Substituting \eref{mean-field-eq-Tr1} and \eref{mean-field-eq-Tr2}
to \eref{mean-field-Tr}, we come to
\begin{equation}\label{mean-field-eq}
    u^{-1} \bar V^z_{00} + \bigl(1 + \eta \bigr) T \Tr \bar G^{z} = 0.
\end{equation}

The mean Green function $\bar G$ is related to the mean spin moment
(per atom) $s_{z}$ by the formula (see \eref{total-spin} in~\ref{Appendix-B})
\begin{equation*}\label{s-g}
     s_{z} = N_\mathrm{a}^{-1} T \Tr \bar G^{z}.
\end{equation*}
Thus, the mean-field equation \eref{mean-field-eq} takes the form
\begin{equation}\label{Vz}
    \bar V_z  = - u \bigl(1 + \eta \bigr) s_z,
\end{equation}
$\bar V_z \equiv N_\mathrm{a}^{-1} \bar V^z_{00}$ being the value of
the mean exchange field. The conservation of the total number of electrons
condition ($\partial \mathcal{F}/ \partial \mu = 0$) yields the equation
on the chemical potential (see~\ref{Appendix-B}):
\begin{equation}\label{mu}
     N_\mathrm{e} = T\Tr\, \bar G.
\end{equation}

\subsection{Equations for the spin fluctuations}
Reduce quadratic form \eref{HK2-decoupled} to a sum of squares using
the momentum-`frequency' representation. By Parseval's identity,
the first term in \eref{HK2-decoupled} can be rewritten as
\begin{equation}\label{Parseval}
   \frac1{U} \sum_{\alpha} \Tr (\Delta V^{\alpha})^2
   = \frac{N_\rmd}{U} \sum_{\mathbf{q}m\alpha} |\Delta V^{\alpha}_{\mathbf{q}m}|^2
   = u^{-1}\sum_{\mathbf{q}m\alpha} |\Delta V^{\alpha}_{\mathbf{q}m}|^2.
\end{equation}
For the trace in the second term of \eref{HK2-decoupled}, using
\eref{V-momentum-frequency} and \eref{G-momentum-frequency} we
have
\begin{eqnarray*}
\fl
    \Tr & \bigl(\bar G \Delta V \bar G \Delta V \bigr) \\
\fl
    & = N_\rmd \sum_{\mathbf{k}\mathbf{k}_1nn_1}
  \sum_{\alpha\beta\gamma_1\gamma_2}
  \bar G^{\gamma_1}_{\mathbf{k}n} \Delta V^{\alpha}_{\mathbf{k-k}_1,\,n-n_1}
  \bar G^{\gamma_2}_{\mathbf{k}_1n_1} \Delta V^{\beta}_{\mathbf{k}_1-\mathbf{k},\,n_1-n}
    \Sp \bigl(\sigma^{\gamma_1} \sigma^{\alpha}
    \sigma^{\gamma_2} \sigma^{\beta}\bigr),
\end{eqnarray*}
where $\gamma_1, \gamma_2 = 0,z$. Calculations show that the summands with $\alpha\neq\beta$
are equal to zero (see~\ref{Appendix-C}). Hence
\begin{equation}\label{HK2-coord}
\fl
   \frac12 \bigl(1 + 3 \eta \bigr)
    T \Tr \bigl( \bar G \Delta V \bar G \Delta V \bigr)
    = \sum_{\mathbf{q}m\alpha} \Delta V^{\alpha}_{\mathbf{q}m}
      \chi^{\alpha}_{\mathbf{q}m} \Delta V^{\alpha}_{\mathbf{-q}-m}
    = \sum_{\mathbf{q}m\alpha} \chi^{\alpha}_{\mathbf{q}m}
    |\Delta V^{\alpha}_{\mathbf{q}m}|^2,
\end{equation}
where $\mathbf{q}=\mathbf{k-k}_1$, $m=n-n_1$, and
\begin{eqnarray}\label{chi}
  \chi^{\alpha}_{\mathbf{q}m}
  &\equiv -\frac12\, \frac{\partial^2 F_1^{(2)}}{\partial
  \bigl(\Delta V^{\alpha}_{\mathbf{q}m}\bigr)
  \partial \bigl(\Delta V^{\alpha}_{-\mathbf{q}-m}\bigr)}
  \nonumber \\
  &= - \frac{N_\rmd}2 \bigl(1 + 3 \eta \bigr) T
   \sum_{\mathbf{k}n} \sum_{\gamma_1\gamma_2}
   \bar G^{\gamma_1}_{\mathbf{k}n} \bar G^{\gamma_2}_{\mathbf{k}-\mathbf{q},\,n-m}
   \Sp \bigl(\sigma^{\gamma_1} \sigma^{\alpha} \sigma^{\gamma_2} \sigma^{\alpha}\bigr)
\end{eqnarray}
is the unenhanced dynamic susceptibility. Substituting \eref{Parseval}
and \eref{HK2-coord} to \eref{HK2-decoupled}, we obtain
\begin{equation}\label{HK-coord}
   F^{(2)}(V)
    = \sum_{\mathbf{q}m\alpha}
      \biggl(u^{-1} - \chi^{\alpha}_{\mathbf{q}m} \biggr)
      |\Delta V^{\alpha}_{\mathbf{q}m}|^2
    \equiv \sum_{\mathbf{q}m\alpha} A^{\alpha}_{\mathbf{q}m}
      |\Delta V^{\alpha}_{\mathbf{q}m}|^2.
    \end{equation}
Thus $\Delta V^{\alpha}_{\mathbf{q}m}$ are statistically
independent Gaussian fluctuations, with the mean squares of
fluctuations
\begin{equation}\label{variance}
 \langle |\Delta V^{\alpha}_{\mathbf{q}m}|^2 \rangle
   =\frac{T}{2A^{\alpha}_{\mathbf{q}m}}
   = \frac{T}{2( u^{-1} - \chi^{\alpha}_{\mathbf{q}m} )}.
\end{equation}

\section{Local approximation of the SFT}
\subsection{Reduction to local fluctuations}
Expressing the mean Green function $\bar G$ in terms of the
chemical  potential $\mu$, mean field $\bar V_z$ and fluctuations
$\langle |\Delta V^{\alpha}_{\mathbf{q}m}|^2 \rangle$, we come to
the closed system of equations \eref{Vz}, \eref{mu} and
\eref{variance}. However, this system of equations is still
excessively difficult for practical computations of average
quantities, like magnetization. Moreover, as a result of the
quadratic approximation \eref{HK-coord} the fluctuations at
different momenta and `frequencies' $\Delta
V^{\alpha}_{\mathbf{q}m}$ become independent, which is an
acceptable approximation only at low temperatures. Therefore,
we proceed with the local approximation of the quadratic form
\eref{HK-coord}:
\begin{eqnarray*}\label{HK-sigle-site}
\fl
   F^{(2)}(V)
   = \sum_{\mathbf{q}m\alpha}
   A^{\alpha}_{\mathbf{q}m} |\Delta V^{\alpha}_{\mathbf{q}m}|^2
   &\approx \sum_{\alpha} A^{\alpha}
   \sum_{\mathbf{q}m} |\Delta V^{\alpha}_{\mathbf{q}m}|^2 \nonumber \\
   &= \sum_{\alpha} N_\mathrm{a} A^{\alpha}
   \biggl(\frac1{N_\mathrm{a}}\sum_{\mathbf{q}m} |\Delta V^{\alpha}_{\mathbf{q}m}|^2\biggr)
   \equiv \sum_{\alpha}  A_{\alpha} \Delta V_{\alpha}^2.
\end{eqnarray*}
Here the coefficient $A_\alpha \equiv N_\mathrm{a} A^\alpha$ is related to
the mean square of the local fluctuation $\langle \Delta V_{\alpha}^2 \rangle$
by the formula
\begin{equation*}\label{variance-single-site}
    A_{\alpha}
    = \frac{T}{2} \langle \Delta V_{\alpha}^2 \rangle^{-1}
    = \frac{T}{2}
    \Biggl( \frac1{N_\mathrm{a}} \sum_{\mathbf{q}m}
    \bigl\langle |\Delta V^{\alpha}_{\mathbf{q}m}|^2 \bigr\rangle \Biggr)^{-1},
\end{equation*}
where
\begin{equation*}\label{average-single-site}
\fl
    \langle \dots \rangle
    = \biggl( \int \exp\Bigl( -\sum_{\alpha} A_{\alpha}
    \Delta V_{\alpha}^2/T \Bigr)\, \rmd \mathbf{V} \biggr)^{-1}
    \int \dots \exp\Bigl( -\sum_{\alpha} A_{\alpha}
    \Delta V_{\alpha}^2/T \Bigr)\, \rmd \mathbf{V}.
\end{equation*}
Taking into account \eref{variance}, for the mean square of
the local fluctuation (`fluctuation', for short) we have
\begin{equation}\label{sum-qm}
    \langle \Delta V_{\alpha}^2 \rangle
    = \frac1{N_\mathrm{a}} \sum_{\mathbf{q}m}
    \bigl\langle |\Delta V^{\alpha}_{\mathbf{q}m}|^2 \bigr\rangle
    = \frac1{N_\mathrm{a}} \sum_{\mathbf{q}m}
    \frac{T}{2( u^{-1} - \chi^{\alpha}_{\mathbf{q}m} )}\,.
\end{equation}

\subsection{Summation over momenta and `frequencies'}
Calculation of the sum \eref{sum-qm} follows \cite{RG97}
and yields essentially the same formulae but with the renormalization
prefactor $1+3\eta$ for the susceptibility. By analytic
continuation from the points $\rmi\omega_m = \rmi 2\pi mT$, summation
over `frequencies' is replaced by the integration over the
energy variable (for details, see~\cite{MR10}):
\begin{equation} \label{Bose-integral}
  \sum_{m}\langle|\Delta V^{\alpha} _{{\bf q}m}|^2\rangle
  = \frac u2 \frac 2\pi \int_0^{\infty} \Bigl(B(\varepsilon) + \frac12 \Bigr)\,
  \mathrm{Im}\, \frac1{1 - u \chi_{\bf q}^{\alpha}(\varepsilon+\rmi0)}\,
  \rmd\varepsilon,
\end{equation}
where $B(\varepsilon)=(\exp(\varepsilon/T) - 1)^{-1}$ is the Bose
function. We discard the temperature-independent term with $1/2$,
assuming that the zero-point fluctuations are already taken into
account in the initial DOS $\nu(\varepsilon)$ calculated by the
density-functional method and in the effective interaction
constant $u$. Using the Tailor expansion
$\chi_{\mathbf{q}}(\varepsilon)\approx \chi_{\mathbf{q}}(0)
+\rmi\varphi_{\mathbf{q}}\varepsilon$ and approximation
\begin{equation}\label{Bose-approx}
 \frac1{\rme^{\varepsilon/T} - 1} \approx \left\{
 \begin{array}{ll}
  T/\varepsilon, \qquad
  & \varepsilon<\varepsilon_0=(\pi^2/6)T, \\
  0, \qquad & \varepsilon>\varepsilon_0,
 \end{array} \right.
\end{equation}
for the Bose function, we come to
\begin{equation} \label{sum-m}
  \sum_{m} \langle|\Delta V^{\alpha} _{{\bf q}m}|^2\rangle
  = \frac{uT}{2\lambda^{\alpha}_\mathbf{q}} \frac2{\pi}
  \arctan \frac{u\varphi^{\alpha}_{\mathbf{q}}\pi^2T}{6\lambda^{\alpha}_\mathbf{q}},
\end{equation}
where
\begin{equation} \label{lambda}
 \lambda^{\alpha}_\mathbf{q}=1-u\chi^{\alpha}_\mathbf{q}(0),
  \qquad
 \varphi_{\mathbf{q}}^\alpha
 = \frac{d \Im\chi^\alpha_{\mathbf{q}}(\varepsilon)}{d\varepsilon}\biggr|_{\varepsilon=0}.
\end{equation}

The approximation \eref{Bose-approx} not only reproduces the behaviour of
the Bose function $B(\varepsilon)$ with respect to thermal energies,
but also has the same first moment
$\int_0^{\infty}\varepsilon B(\varepsilon)\, \rmd\varepsilon = (\pi T)^2/6$,
which essentially defines the upper bound $\varepsilon_0$. Thus,
the approximation \eref{Bose-approx} is well justified. Its another advantage
is the possibility of the straight-forward proceeding to the static limit
at high temperatures, when the argument of the arctangent in \eref{sum-m}
is much larger than unity.

The function $\lambda^{\alpha}_\mathbf{q}$ is calculated using the
expansion for the static susceptibility:
\begin{equation} \label{chi(0)}
 \chi^{\alpha}_\mathbf{q}(0)
 \approx \chi^{\alpha}_{0}(0) + B^\alpha q^2,
\end{equation}
where $\chi^{\alpha}_0(0)$ is the static uniform susceptibility,
and coefficient $B^\alpha$ is obtained from the local
susceptibility $\chi^\alpha_\mathrm{L}(0) = N_\mathrm{a}^{-1} \sum_\mathbf{q}
\chi^\alpha_\mathbf{q}(0)$. Substituting \eref{chi(0)} in the
first equality \eref{lambda}, we get
\begin{equation} \label{lambda-q}
\lambda^{\alpha}_\mathbf{q} = \lambda^{\alpha}_0 +
(\lambda^{\alpha}_\mathrm{L} - \lambda^{\alpha}_0)q^2/\overline{q^2},
\end{equation}
where $\lambda^{\alpha}_0 = 1 - u\chi^{\alpha}_0(0)$,
$\lambda^{\alpha}_\mathrm{L} = 1 - u\chi^{\alpha}_\mathrm{L}(0)$, and
$\overline{q^2} = N_\mathrm{a}^{-1} \sum_\mathbf{q} q^2$. The function
$\varphi^\alpha_{\mathbf{q}}$ is approximated by its mean value:
\begin{equation} \label{varphi}
\varphi^\alpha_{\mathbf{q}} \approx N_\mathrm{a}^{-1} \sum_\mathbf{q}
\varphi^\alpha_{\mathbf{q}} = \varphi^\alpha_\mathrm{L}.
\end{equation}

The summation over $\mathbf{q}$ is carried out by the integration
over the Brillouin zone, approximated for simplicity by the sphere
of the same volume. Using \eref{sum-m}, \eref{lambda-q} and
\eref{varphi}, for the local fluctuation \eref{sum-qm} we
finally obtain
\begin{equation}\label{zeta-alpha-2}
 \zeta^\alpha
 \equiv \langle \Delta V_{\alpha}^2 \rangle
 = \frac{uT}{2\lambda^{\alpha}_\mathrm{L}}\int_0^1 \frac1{a^2_{\alpha} + b^2_{\alpha}k^2}
 \frac{2}{\pi} \arctan\frac{c_{\alpha}}{a^2_{\alpha}+b^2_{\alpha}k^2}3k^2\, \rmd k,
\end{equation}
where
\begin{equation*}
  a^2_{\alpha} = \lambda^{\alpha}_0/ \lambda^{\alpha}_\mathrm{L},
  \quad
  b^2_{\alpha} = (1 - a^2_{\alpha})/0.6,
  \quad
  c_{\alpha} = u\varphi^{\alpha}_\mathrm{L}\pi^2 T/(6\lambda^{\alpha}_\mathrm{L}).
\end{equation*}
Relation \eref{zeta-alpha-2} is obtained using the integration over
the Brillouin zone with the Bose distribution and a simple dispersion
relation, whose parameters are chosen to give a correct value of
the local susceptibility. Thus, expression \eref{zeta-alpha-2} for
the mean square of the spin fluctuations is self-consistent
and does not contain any free parameters.

\subsection{Mean single-site Green function}
To calculate the local susceptibility
$\chi^{\alpha}_\mathrm{L}(\varepsilon)=\chi^{\alpha}_\mathrm{L}(0)
+\rmi\varphi^{\alpha}_{L}\varepsilon$,
we replace the mean Green function $\bar G$ in \eref{chi} by its
site-diagonal part:
\begin{equation}\label{Green-singlesite}
  \bar G_{jj'n} = \bar G_{j-j',\,n}
  \approx \bar G_{0n} \delta_{jj'} \equiv g_{n} \delta_{jj'},
\end{equation}
so that the Fourier transform is $\mathbf{k}$-independent: $\bar
G_{\mathbf{k}n}=g_{n}$. Thus, we rewrite \eref{chi} as
\begin{eqnarray*}\label{chi-singlesite}
\fl
  \chi^{\alpha}_{L} (\rmi\omega_m)
   = - \frac{N_\rmd}2 \bigl(1 + 3 \eta \bigr)
   T \sum_{n} \sum_{\gamma_1\gamma_2}
   g^{\gamma_1}(\rmi\omega_n) g^{\gamma_2}(\rmi\omega_n - \rmi\omega_m)
   \Sp \bigl(\sigma^{\gamma_1} \sigma^{\alpha}
   \sigma^{\gamma_2} \sigma^{\alpha}\bigr).
\end{eqnarray*}
Using analytic continuation first from points $\rmi\omega_n$ and then
from $\rmi\omega_m$, we replace the sum over `frequencies' by the
integral over the energy variable:
\begin{eqnarray*}\label{chi-integral}
\fl
  \chi^{\alpha}_{L} (z)
   = - \frac{N_\rmd}{2\pi} \bigl(1 + 3 \eta \bigr) \sum_{\gamma_1\gamma_2}
   \int \Im \Bigl( g^{\gamma_1}(\varepsilon)
   &[g^{\gamma_2}(\varepsilon-z) \nonumber \\
   &+ g^{\gamma_2}(\varepsilon+z)] \Sp \bigl(\sigma^{\gamma_1}
   \sigma^{\alpha} \sigma^{\gamma_2}
   \sigma^{\alpha}\bigr) \Bigr) f(\varepsilon)\, \rmd\varepsilon,
\end{eqnarray*}
where $f(\varepsilon)=[\exp((\varepsilon-\mu)/T)+1]^{-1}$ is the
Fermi function, and $g(\varepsilon) = g(\varepsilon-\rmi0)$.
The $2\times2$ matrix $g(\varepsilon)$ is spin-diagonal:
$g^x(\varepsilon)=g^y(\varepsilon)=0$, same as $\bar
G_{\mathbf{k}n}$ (see \eref{G-xy}). Denoting the diagonal elements
of $g(\varepsilon)$ by $g_\uparrow(\varepsilon)$ and
$g_\downarrow(\varepsilon)$, we have
$g^0(\varepsilon)=\frac12[g_\uparrow(\varepsilon) +
g_\downarrow(\varepsilon)]$ and
$g^z(\varepsilon)=\frac12[g_\uparrow(\varepsilon) -
g_\downarrow(\varepsilon)]$. Hence\footnote{Recall that
$\chi^x_{\mathbf{q}m}=\chi^y_{\mathbf{q}m}$ due to axial
symmetry (for details see~\ref{Appendix-C}).}
\begin{eqnarray}\label{sft-13}
 \chi^x_\mathrm{L}(0) &= -\frac{N_\rmd(1+3\eta)}{\pi}
 \int\Im(g_{\uparrow} g_{\downarrow})\,f\, \rmd\varepsilon, \\
 \varphi^x_\mathrm{L} &= \frac{N_\rmd(1+3\eta)}{\pi}
 \int\Im g_{\uparrow}\,\Im g_{\downarrow}
 \left(-\frac{\partial f}{\partial\varepsilon} \right)
 \rmd\varepsilon, \\
 \chi^z_\mathrm{L}(0)
 &= -\frac{N_\rmd(1+3\eta)}{2\pi} \int\bigl(\Im g^2_{\uparrow}
 + \Im g^2_{\downarrow}\bigr)\,f\, \rmd\varepsilon, \\
 \varphi^z_\mathrm{L}
 &= \frac{N_\rmd(1+3\eta)}{2\pi} \int\Bigl[\bigl(\Im g_{\uparrow}\bigr)^2
 + \bigl(\Im g_{\downarrow}\bigr)^2\Bigr]
 \left(-\frac{\partial f}{\partial\varepsilon} \right) \rmd\varepsilon.
 \label{sft-13a}
\end{eqnarray}

Similarly, in equations for the mean field \eref{Vz} and chemical potential
\eref{mu}, we use the mean single-site Green function \eref{Green-singlesite}
and replace the sum over `frequencies' by the integral over the energy
variable. Thus, we rewrite equation \eref{Vz} as
\begin{equation}\label{Vz-singlesite}
     \bar V_z
     = -u(1+\eta) \frac{N_\rmd}{2\pi} \int \Im \bigl( g_{\uparrow} - g_{\downarrow}
     \bigr) f\, \rmd\varepsilon
\end{equation}
and equation \eref{mu} as
\begin{equation}\label{mu-coord}
  N_\mathrm{e} = N_\mathrm{a} \frac{N_\rmd}{\pi} \int \Im \bigl( g_{\uparrow}
      + g_{\downarrow} \bigr) f \, \rmd\varepsilon.
\end{equation}

In formulae \eref{sft-13}--\eref{mu-coord}, the mean single-site
Green function is given by
\begin{equation*}
 g_{\sigma}(\varepsilon) = \int \frac{\nu(\varepsilon')}
 {\varepsilon - \sigma \bar V_z - \Delta\Sigma_{\sigma}(\varepsilon) - \varepsilon'}\,
 \rmd\varepsilon',
  \label{sft-11}
\end{equation*}
where $\sigma = \uparrow,\downarrow$ or $\pm1$ is the spin index,
${\nu(\varepsilon)}$ is the non-magnetic DOS, and
$\Delta\Sigma_{\sigma}(\varepsilon)$ is the fluctuational contribution
to the self-energy part, calculated by the formula
\begin{equation*}
 \Delta\Sigma_{\sigma}(\varepsilon)
 = \frac{g_{\sigma}(\varepsilon)\zeta^z}{1+2\sigma \bar V_z g_{\sigma}(\varepsilon)}
 + 2g_{\bar\sigma}(\varepsilon)\zeta^x,
 \qquad
 \bar\sigma\equiv{-\sigma}.
\end{equation*}
The latter is obtained from the coherent potential equation
\[
    \Delta \Sigma = \langle \Delta V [1 - g(\Delta V - \Delta \Sigma)]^{-1} \rangle
\]
in the second order with respect to the fluctuations $\Delta V$ \cite{RG97}.

\subsection{Correction coefficient}
Final computational formulae of the extended SFT differ from the
ones in \cite{RG97} by the renormalization of susceptibility
\eref{sft-13}--\eref{sft-13a} and the mean field
\eref{Vz-singlesite} that depend on the coefficient $\eta$ defined
in \eref{eta-def} (with $\eta=0$ the system of equations reduces to
the one in~\cite{RG97}). Using the single-site and quasi-static
approximations, we come to the following
expression for $\eta$ (see~\ref{Appendix-D}):
\begin{equation}\label{a-coeff-value}
\fl
   \eta \approx \frac1{\pi T} \biggl[
   2\zeta^x \!\!\!
   \int \bigl( \Re g_{\uparrow} \Im g_{\downarrow}
   + \Re g_{\downarrow} \Im g_{\uparrow} \bigr) f\, \rmd\varepsilon \,
   + \zeta^z \!\!\!
   \int \bigl( \Re g_{\uparrow} \Im g_{\uparrow}
   + \Re g_{\downarrow} \Im g_{\downarrow} \bigr) f\,
   \rmd\varepsilon \biggr].
\end{equation}

The impact of the corrections due to the third- and fourth-order
terms becomes critical at high temperatures. Therefore,
coefficient $\eta$ can be estimated by the reduced formula
\begin{equation}\label{eta-4}
   \eta \approx \frac{6 \bar \zeta}{\pi T} \,
   \int \Re g^0 \Im g^0 f\, \rmd\varepsilon
   \equiv \frac{c}{T} \bar \zeta,
\end{equation}
where $\bar \zeta = (2\zeta^{x} + \zeta^{z})/3$ is the mean
fluctuation, and $g^0 = (g_\uparrow + g_\downarrow)/2$. In the
ferromagnetic region, formula \eref{eta-4} follows from the
initial formula \eref{a-coeff-value} in the approximation
$g_{\uparrow} = g_{\downarrow}$. In the paramagnetic region,
formulae \eref{a-coeff-value} and \eref{eta-4} coincide.

\section{Numerical results}
\begin{figure}[h]
\begin{center}
\centerline{\includegraphics[scale=0.92]{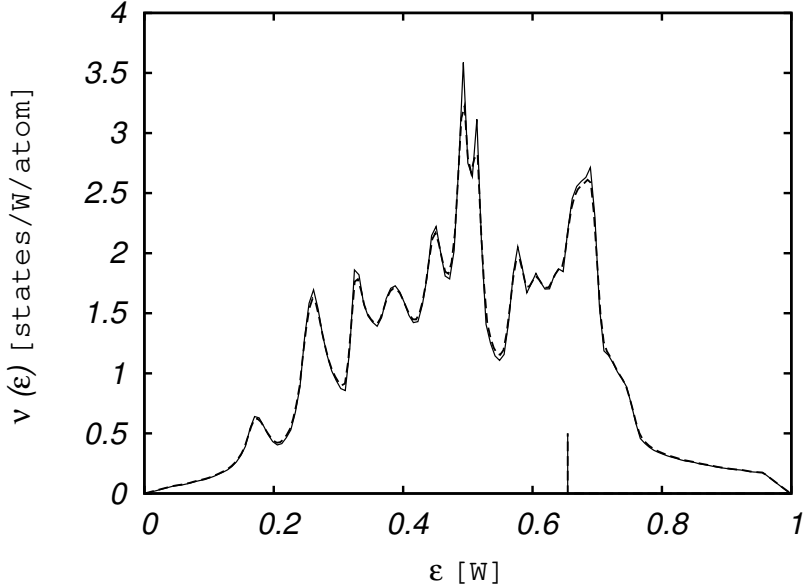}}
\caption{The DOS of the d band of non-magnetic
$\mbox{Fe}_{0.65}\mbox{Ni}_{0.35}$, obtained from \cite{JPS85}
(\full), and the one smoothed out by convolution with the Lorentz
function of the half-width $\Gamma=0.001$ (\broken). The energy
$\varepsilon$ and half-width $\Gamma$ are in units of the
bandwidth $\mathrm{W}=9.70$ eV. The vertical line indicates the position of
the Fermi level $\varepsilon_F$.}
\label{DOS}
\end{center}
\end{figure}

The extended SFT is investigated by the example of the Invar alloy
$\mbox{Fe}_{0.65}\mbox{Ni}_{0.35}$. The initial non-magnetic DOS
$\nu(\varepsilon)$ (see figure~\ref{DOS}) is formed from
the two spin-polarized densities obtained from
the self-consistent calculation for the completely disordered
$\mbox{Fe}_{0.65}\mbox{Ni}_{0.35}$ alloy~\cite{JPS85}.
The experimental value of the spin magnetic moment per atom
$m_0^\mathrm{exp} = 1.70\,\mu_\mathrm{B}$
(where $\mu_\mathrm{B}$ is the Bohr magneton),
used to determine the effective interaction constant $u$,
is taken from~\cite{CH63}.

Note that we neglect here the fine effects of atomic and/or magnetic
short-range order (see, e.g.,~\cite{RKM07, CEE02} and references therein).
Moreover, the magnetic moment $m_0^\mathrm{exp}$ and DOS $\nu(\varepsilon)$
represent the values per \emph{averaged} atom. However, as stated in
the introduction, even with these initial data one can calculate
the temperature dependence of magnetic properties of an alloy in the SFT.

\begin{figure}[h]
\begin{center}
\centerline{\includegraphics[scale=0.92]{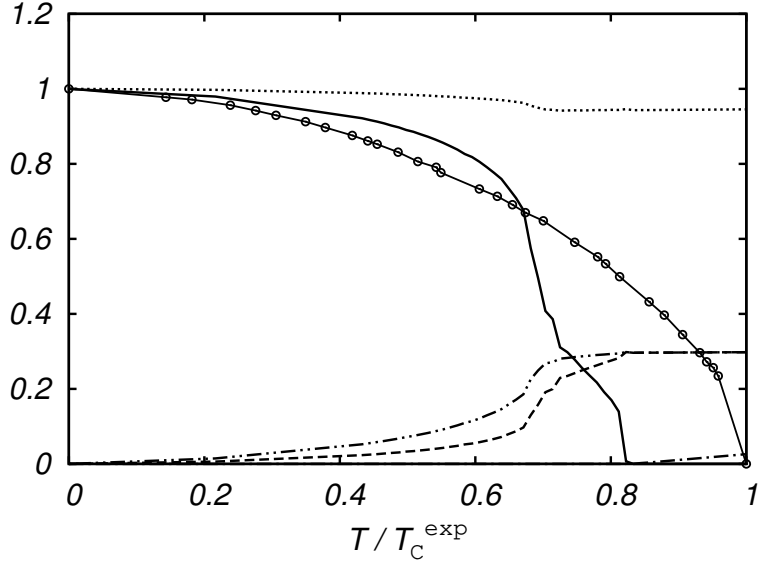}}
\caption{The magnetization $m(T)/m(0)$ (\full\, calculation,
$\circ\!\circ\!\circ\circ$ experiment~\cite{CH63}),
the mean square of spin fluctuations $\zeta^x$ (\dashddot) and
$\zeta^z$ (\broken) in units of $um(0)/(2\mu_{\mathrm{B}})$,
the reciprocal paramagnetic susceptibility $\chi^{-1}(T)$
(\chain) in units of $T^{\mathrm{exp}}_{\mathrm{C}}/\mu^2_{\mathrm{B}}$,
and the local magnetic moment $m_{\mathrm{L}}(T)/m(0)$ ($\dotted$) of the Invar
$\mbox{Fe}_{0.65}\mbox{Ni}_{0.35}$ calculated in the Gaussian
SFT as functions of the reduced temperature
$T/T_{\mathrm{C}}^{\mathrm{exp}}$.}
\label{DNA-results}
\end{center}
\end{figure}

\Fref{DNA-results} presents the basic magnetic characteristics for the
$\mbox{Fe}_{0.65}\mbox{Ni}_{0.35}$ Invar calculated within
the Gaussian SFT \cite{RG97}. Clearly, at high temperatures,
the calculated magnetization curve does not fit well enough
the experimental one. For the Curie temperature, we obtain
$T_{\mathrm{C}}=0.83T_{\mathrm{C}}^{\mathrm{exp}}$. But most important,
the calculated curve $m(T)$ has the inflection
(see the discussion in~\cite{RM08}).

In~\cite{RMG09} we took into account the higher terms in the expansion of
the free energy $F(V)$ by using the simplified expression \eref{eta-4} for
the correction coefficient $\eta$ with $c=-0.015\mathrm{W}^{-1}$
($\mathrm{W}=9.70$ eV is the bandwidth). The calculation gave nearly
full agreement with experiment for the Curie temperature:
$T_{\mathrm{C}}=1.02T_{\mathrm{C}}^{\mathrm{exp}}$
($T_{\mathrm{C}}^{\mathrm{exp}}=520 \,\mathrm{K}$ \cite{CH63}),
for the paramagnetic Curie point:
$\Theta_{\mathrm{C}}=1.06T_{\mathrm{C}}^{\mathrm{exp}}$,
for the effective magnetic moment:
$m_{\mathrm{eff}}=0.90m_{\mathrm{eff}}^{\mathrm{exp}}$
($m_{\mathrm{eff}}^{\mathrm{exp}}=3.3 \, \mu_\mathrm{B}$ \cite{MAC80})
and for the local magnetic moment $m_\mathrm{L}(T)$
(see the discussion in~\cite{Res04a}).
As can be seen from figure~1 in~\cite{RMG09},
a sharp increase of the fluctuations and sharp decrease
in magnetization at high temperatures, which occurred in~\cite{RM08},
disappear in the extended SFT.

\begin{figure}[h]
\begin{center}
\centerline{\includegraphics[scale=0.92]{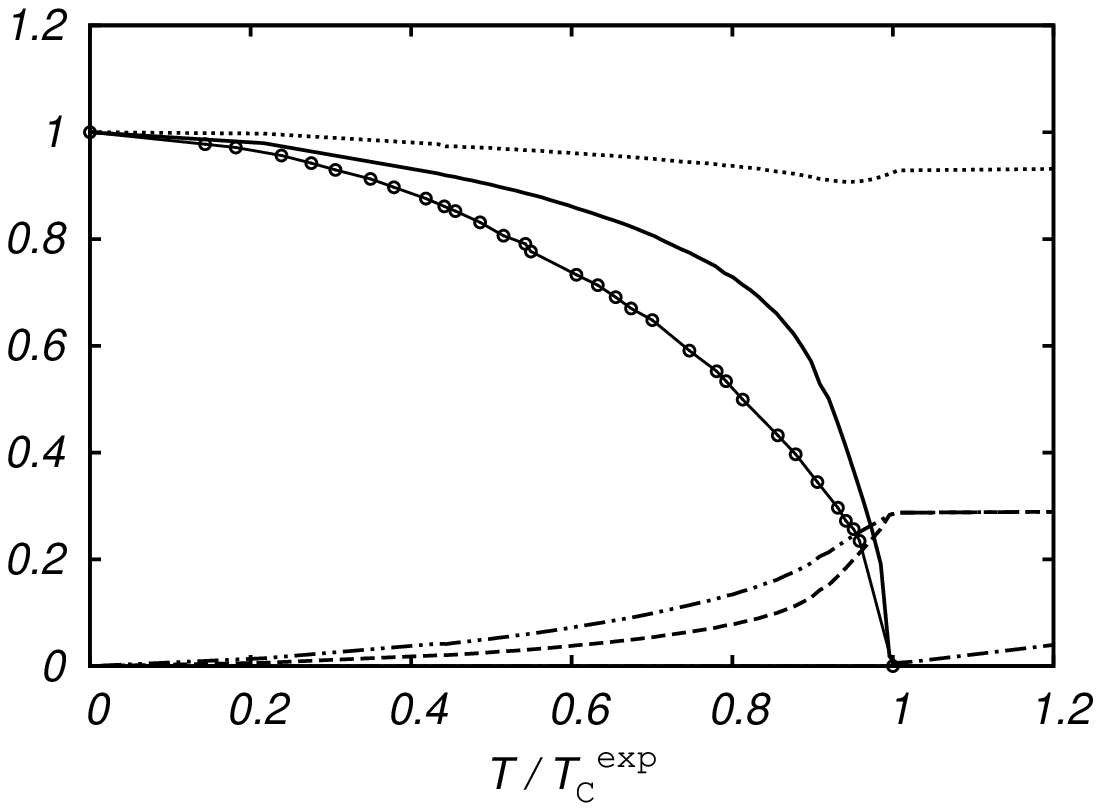}}
\caption{As \fref{DNA-results}, but calculated in the extended SFT
with the expression~\eref{a-coeff-value} for $\eta$.}
\label{extended SFT-results}
\end{center}
\end{figure}

On the whole, the curve for the magnetization in \cite{RMG09} fits
the experimental one well enough. However, the inflection in
the temperature dependence, reported in~\cite{RM08}, does not vanish entirely.
Therefore, in the present paper, we apply the expression~\eref{a-coeff-value},
which alternates with $T$ in a self-consistent way. This finally gives
a smooth curve without the inflection (\fref{extended SFT-results}).

A qualitative analysis of our equations explains the mechanism that leads to
a jump transition of magnetization in the simple Gaussian theory and
its elimination in the extended SFT. Indeed, in the presence of the external
magnetic field $h$ the mean-field equation \eref{Vz} takes the form
\begin{equation*}
   \bar V_z(h)  = - u \bigl(1 + \eta \bigr) s_z(\bar V_z(h) + h).
\end{equation*}
Hence for the enhanced magnetic susceptibility we get
\begin{equation}\label{magnetic-suscept}
    - \frac{\rmd s_z}{\rmd h} = \frac{\chi_0}{1-u(1+\eta)\chi_0},
\end{equation}
where $\chi_0=-\partial s_z/\partial h$ is the unenhanced
(with constant $\bar V_z$) susceptibility of non-interacting electrons.
The fluctuation \eref{zeta-alpha-2} is proportional to
\begin{equation}\label{magnetic-fluct}
    \langle \Delta V_{\alpha}^2 \rangle \propto \frac{1}{1-u(1+3\eta)\chi_0}.
\end{equation}
The enhanced susceptibility \eref{magnetic-suscept} diverges at the critical
temperature (the condition of the transition from ferro- to paramagnetic state).
If, at the same time, the fluctuations also increase sharply, i.e.
the derivative tends to infinity, then there exist an extra (unstable)
solution and hence a jump transition to the paramagnetic state.
Such a scenario takes place in the Gaussian approach ($\eta=0$),
where the susceptibility \eref{magnetic-suscept} and the amplitude of
the fluctuations \eref{magnetic-fluct} diverge simultaneously.
In the present variant of the theory ($\eta<0$) the fluctuations
are weakened, and as a result we observe a continuous magnetic transition.

There are different ways to go beyond the approximation of
the non-interacting spin fluctuations. In this respect our approach
can be treated as a variant of the SCR theory. Formulae
\eref{magnetic-suscept} and \eref{magnetic-fluct}, together with \eref{eta-4},
schematically demonstrate the character of our renormalizations.

\section{Conclusions}
We have developed an approximation to the SFT that describes
the thermodynamics of magnetic characteristics. Our approach
takes into account both dynamics and non-locality of thermal
spin fluctuations, as well as their mode-mode 'frequency'
interactions. As the initial data, the calculation employs
the value of the magnetic moment and \emph{ab initio} DOS,
calculated at zero temperature. Further self-consistent treatment
of thermal fluctuations, including 'large' non-Gaussian fluctuations,
makes the approximation of the SFT applicable for all temperatures.
Particularly, the present theory yields a proper second-order phase
transition from ferro- to paramagnetic state.

\ack This work was supported in part by the Russian Foundation for
Basic Research (grant nos. 08-01-00685 and 08-02-00904) and by the
Ministry of Education and Science of the Russian Federation (grant
no. 2.1.1/2000).

\appendix
\section{Relation between the free energy and Green function}
\label{Appendix-A} In this section, we derive the relation between
the free energy of electrons in the external field and the Green
function used in the main text\footnote{The relation between
the thermodynamic potential $\Omega_1(V)$ and Green function
$G(V)$, in the case of an \emph{arbitrary} perturbation $V$,
was obtained in~\cite{AGD63}. Without a proof the formula,
analogous to \eref{Omega1_GV}, was presented in~\cite{Kak92}.}.
Exact expression for the constant term of the free energy allows us
to obtain the equation on the chemical potential as the conservation
of number of particles condition in the next Appendix.

The thermodynamic potential of non-interacting electrons in the
external field $V$ is defined as
\begin{equation}\label{F(V)-def}
   \Omega_1(V) = - T \ln\Tr
   \biggl[T_{\tau} \exp\biggl(- \int_{0}^{1/T} H'(V)\, \rmd\tau\biggr) \biggr],
\end{equation}
where the Hamiltonian corresponding to the grand canonical ensemble:
\begin{equation*}
  H'(V) = H_0' + V(\tau),
\end{equation*}
consists of the Hamiltonian $H_0'=H_0-\mu N_\mathrm{e}$ and the external field
\begin{equation}\label{external-field}
 V(\tau)
  = \sum_{j\sigma\sigma'}
  V_{j\sigma\sigma'}(\tau) a^+_{j\sigma}(\tau) a_{j\sigma'}(\tau)
  = \sum_{j} \Sp \bigl( V_{j}(\tau) \rho_{j}(\tau) \bigr),
\end{equation}
the interaction representation of an operator $\mathcal{O}$ is defined as
\begin{equation*}
  {\mathcal O}(\tau) = \rme^{\tau H'_0} {\mathcal O} \rme^{-\tau H'_0},
\end{equation*}
$T_\tau$ stands for the `time'-ordering operator,
and $\Tr$ contains the summation over states with any number of particles.
Note that in $H'(V)$, $\mu$ is the chemical potential,
$N_\mathrm{e}=\sum_{j\sigma} a^+_{j\sigma}(\tau) a_{j\sigma}(\tau)$
is the number of particles operator, $a^+_{j\sigma}$ and $a_{j\sigma}$
are the creation/annihilation operators for Wannier states,
and $\rho_j$ is the local spin density matrix, with elements
\begin{equation}\label{rho-def}
  \rho_{j\sigma\sigma'} =  a^+_{j\sigma'}  a_{j\sigma}.
\end{equation}

The method to relate the thermodynamic potential $\Omega_1(V)$
to the Green function
\begin{equation}\label{G-H'}
  G(V) = \bigl(z - H'(V)\bigr)^{-1}
\end{equation}
is to vary the strength of the external field from $0$ to $V$.
To this end, we consider
\begin{equation}\label{H_lambda}
  H'(\lambda) = H'_0 + \lambda V,
\end{equation}
so that $H'(0) = H'_0$ and $H'(1) = H'(V)$. The thermodynamic potential
corresponding to $H'(\lambda)$ is
\begin{equation*}
   \Omega_1(\lambda) = - T \ln\Tr \biggl[T_{\tau}
   \exp\biggl(- \int_{0}^{1/T} H'(\lambda)\, \rmd\tau\biggr) \biggr].
\end{equation*}
Using formula \eref{H_lambda}, we find the derivative of $\Omega_1(\lambda)$
with respect to $\lambda$ (cf.~\cite{AGD63}):
\begin{equation}\label{d_lambda_raw}
    \frac{\partial \Omega_1(\lambda)}{\partial \lambda}
    = \biggl\langle T \int_{0}^{1/T} V(\tau)\, \rmd\tau \biggr\rangle_{\lambda},
\end{equation}
where the averaging $\langle \dots \rangle_\lambda$ of an arbitrary operator ${\mathcal{O}}$
is defined as
\begin{equation}\label{averaging-def}
    \bigl\langle {\mathcal{O}}\, \bigr\rangle_{\lambda} \equiv
    \frac{ \Tr \Bigl( {\mathcal{O}}\, T_{\tau}
    \exp\bigl(-\int_0^{1/T} H'(\lambda)\, \rmd\tau\bigr) \Bigr)}{
    \Tr \Bigl(T_{\tau} \exp\bigl(-\int_0^{1/T} H'(\lambda)\,
    \rmd\tau\bigr) \Bigr)}.
\end{equation}

General formula \eref{d_lambda_raw} holds for any perturbation $V$,
not necessarily one-particle operator. For a non-interacting system,
substituting \eref{external-field} for $V$ in \eref{d_lambda_raw}
and rearranging, we write
\begin{equation}\label{d_lambda}
    \frac{\partial \Omega_1(\lambda)}{\partial \lambda}
    = T \int\limits_0^{1/T} \sum_j \Sp
    \Bigl( V_j(\tau) \bigl\langle \rho_j(\tau )
    \bigr\rangle_{\lambda} \Bigr)\, \rmd\tau.
\end{equation}
The average of the spin density matrix in the interaction
representation is related to the Green function:
\begin{equation}\label{rho-G-lambda}
    \bigl\langle \rho_{j\sigma\sigma'}(\tau) \bigr\rangle_{\lambda}
    = - \bigl\langle T_{\tau} a_{j\sigma}(\tau)\, a_{j\sigma'}^+(\tau+0)
    \bigr\rangle_{\lambda}
    \equiv G^{\lambda}_{j\sigma\sigma'}(\tau,\tau).
\end{equation}
Substituting \eref{rho-G-lambda} to \eref{d_lambda}, we come to
\begin{equation*}
   \frac{\partial \Omega_1(\lambda)}{\partial \lambda}
   = T \Tr (VG^{\lambda}).
\end{equation*}
Integration over $\lambda$ between $0$ and $1$ yields
\begin{equation}\label{lambda-int-01}
    \Omega_1(1) -  \Omega_1(0)
    = \int_0^1 T \Tr (V G^{\lambda})\, \rmd\lambda,
\end{equation}
where $\Omega_1(1)=\Omega_1(V)$ and $\Omega_1(0)
= - T \ln \mathrm{Tr}\, \exp(-H'_0/T)$.

The Green function $G^{\lambda}$ of non-interacting electrons satisfies the equation
\begin{equation}\label{Dyson-lambda}
   G^{\lambda} = G_0 + \lambda G_0 V G^{\lambda},
\end{equation}
where $G_0$ corresponds to $H_0$. Express $G^{\lambda}$ from
equation \eref{Dyson-lambda} and substitute to the right-hand side of
\eref{lambda-int-01}. Using the cyclic property of trace, we get
\begin{eqnarray*}
\fl
    \quad \Omega_1(1) -  \Omega_1(0)
    &= T \Tr \int_0^1 (1 - \lambda G_0 V)^{-1} G_0 V\, \rmd\lambda \\
    &= - T \Tr \int_0^1 \frac{\rmd}{\rmd\lambda} \ln (1 - \lambda G_0 V)\,
           \rmd\lambda
     = - T \Tr \ln (1 - G_0 V).
\end{eqnarray*}
Using equation~\eref{Dyson-lambda} and the fact that $G^{1}=G(V)$, we
come to\footnote{Recall the formula $\Tr\ln(A B) = \Tr\ln A + \Tr\ln B$,
which is valid for arbitrary matrices $A$ and $B$.}
\begin{equation}\label{Omega1_GV}
    \Omega_1(V) = - T \ln \mathrm{Tr}\,
    \exp(-H'_0/T) -T \Tr \ln G_0 + T \Tr \ln G(V).
\end{equation}

The free energy $F_1(V)$ is related to the thermodynamic potential
$\Omega_1(V)$ by the formula
\[
 F_1(V) = - T \ln\Tr \biggl[T_{\tau}
              \exp\biggl(- \int_{0}^{1/T} H(V)\, \rmd\tau\biggr) \biggr]
        = \Omega_1(V) + \mu N_\mathrm{e},
\]
where $H(V)=H_0+V$ is the Hamiltonian corresponding to the canonical ensemble,
and the number of electrons $N_\mathrm{e}$ is fixed. Hence formula
\eref{Omega1_GV} can be rewritten as
\begin{equation}\label{F_GV}
    F_1(V) = - T \ln \mathrm{Tr}\, \exp(-H_0/T) -T \Tr \ln G_0 + T \Tr \ln G(V).
\end{equation}
Omitting the first and second terms, that do not depend on $V$, we come to
formula \eref{F1-def} of the main text. The matrix of the Green function
\eref{G-H'} for the system with $N_\mathrm{e}$ electrons reduces to
(see formula~\eref{G-def} in the main text)
\begin{equation}\label{G-append}
  G(V) = (z + \mu - H_0 - V)^{-1}.
\end{equation}

\section{Formulae for the total charge and spin moment}
\label{Appendix-B}
In this section, we express the mean spin moment and total number of electrons
in terms of the mean Green function.

The Green function $G(V)$ is related to the spin density matrix $\rho$
by formula \eref{rho-G-lambda}:
\[
  \langle \rho (\tau) \rangle_V = G(\tau,\tau),
\]
where we write $\langle \dots  \rangle_V$ instead of $\langle \dots \rangle_1$
defined by \eref{averaging-def}. As any Hermitian $2\times2$ matrix,
the local spin density matrix \eref{rho-def} can be expressed as
\begin{equation*}
  \rho_j = \rho^0_j \sigma^0
  + \mbox{\boldmath$\rho$}_j \mbox{\boldmath$\sigma$}
  = \sum_{\mu} \rho^{\mu}_j \sigma^{\mu},
\end{equation*}
with the coefficients
\begin{equation}
   \rho_j^{\mu} = \frac12\mathrm{Sp}\,(\sigma^{\mu} \rho_j),
  \quad \quad \mu=0,x,y,z.
 \label{rho-Pauli-2}
\end{equation}
Here $\sigma^0$ is the $2\times2$ unity matrix, and $\sigma^{\alpha}$
($\alpha=x,y,z$) are the Pauli matrices. Formulae \eref{rho-def} and
(\ref{rho-Pauli-2}) yield that the scalar component $\rho_j^{0}$
is equal to one half of the local charge:
\begin{equation}\label{rho-charge}
  \rho_j^0
  = \frac12( a^+_{j\uparrow}  a_{j\uparrow} +  a^+_{j\downarrow}  a_{j\downarrow})
  = \frac12( n_{j\uparrow} +  n_{j\downarrow}) = \frac12  n_j,
\end{equation}
and the vector component $\mbox{\boldmath$\rho$}_j$ is equal to
the local spin $\mbox{\boldmath$ s$}_j=(s^{x}_j,s^{y}_j,s^{z}_j)$.
Indeed, writing the operator $\mbox{\boldmath$ s$}_j$ in
the second-quantisation form, from \eref{rho-def} we find
\[
   s^{\alpha}_j
  = \sum_{\sigma\sigma'} s^{\alpha}_{\sigma\sigma'}
   a^+_{j\sigma}  a_{j\sigma'}
  = \sum_{\sigma\sigma'} s^{\alpha}_{\sigma\sigma'}
   \rho_{j\sigma'\sigma}
  = \mathrm{Sp}\,(s^{\alpha} \rho_j).
\]
Since the spin operator can be represented by the Pauli matrices:
$\mbox{\boldmath$ s$}=\frac12\mbox{\boldmath$ \sigma$}$,
using \eref{rho-Pauli-2} we obtain
\begin{equation}\label{rho-spin}
 s^{\alpha}_j
  = \frac12\mathrm{Sp}\,(\sigma^{\alpha} \rho_j)
  = \rho_j^{\alpha}.
\end{equation}

Relations \eref{rho-def}, \eref{rho-charge} and \eref{rho-spin} lead to
\begin{equation*}
    \frac12 \langle \langle n_j (\tau) \rangle_V \rangle
     = \bar G_j^{0}(\tau,\tau),
    \qquad
    \langle \langle s_j^{\alpha} (\tau) \rangle_V \rangle
     = \bar G_j^{\alpha}(\tau,\tau),
\end{equation*}
where $\langle \dots \rangle$ is the averaging over all configurations of
the field $V$ with the probability density $p(V)\propto\exp(-\Omega(V)/T)$.
Introduce the Fourier transform
\[
  G_j(\tau,\tau')
  = T \sum_{nn'} G_j(\rmi\omega_n,\rmi\omega_n') \rme^{-\rmi \omega_n \tau}
  \rme^{\rmi \omega_n' \tau'}.
\]
Using the `time'-invariance of the mean Green function, we have
\[
  \bar G_{j}(\tau,\tau) = T \sum_n \bar G_j (\rmi\omega_n)
  \equiv T \sum_n \bar G_{jn}.
\]
Hence, summing over the sites and bands, we come to the expressions for
the total number of electrons:
\begin{equation*}
    N_\mathrm{e} = N_\rmd \sum_{j} \langle \langle n_j \rangle_V \rangle
    = 2 N_\rmd T \sum_{jn} \bar G^0_{jn} = N_\rmd T \sum_{jn} \Sp \bar G_{jn}
    = T \Tr \bar G
\end{equation*}
and the total $z$-projection of spin moment:
\begin{equation}\label{total-spin}
    S^{z} = N_\rmd \sum_{j} \langle \langle s^{z}_j \rangle_V \rangle
    = T \Tr \bar G^z,
\end{equation}
$S_x$ and $S_y$ being equal to zero in the ferromagnetic case, since $\bar G^{x}=0$
and $\bar G^{y}=0$. Analogously, for the non-interacting electrons we have
\begin{equation}\label{charge-G0}
  N_\mathrm{e} = T \Tr G_0.
\end{equation}

Now consider the total free energy $\mathcal{F} = - T \ln Z$, where
the partition function $Z$ is defined by \eref{Z}. Since $\mu$ is
the Lagrange multiplier in the expression for the free energy:
$\mathcal{F}=\Omega+\mu N_\mathrm{e}$, the equation on $\mu$
follows from the extremum condition:
\begin{equation*}
 \frac{\partial \mathcal{F}}{\partial \mu} = 0.
\end{equation*}
The latter is exactly the conservation of the electrons condition.
Since $F_0(V)$ does not depend on $\mu$, differentiation of
$\mathcal{F}$ with respect to $\mu$ yields
\begin{eqnarray*}
\fl
    \frac{\partial \mathcal{F}}{\partial \mu}
    = \frac1{Z} \left(\frac{\displaystyle
    \int \rme^{-F_0(V)/T}\, \rme^{-F_1(V)/T}\,
    \frac{\partial F_1(V)}{\partial \mu}\, DV(\tau)}{\displaystyle
    \int \rme^{-F_0(V)/T}\, DV(\tau)} \right)
    \equiv \left\langle\frac{\partial F_1(V)}{\partial \mu}\right\rangle = 0.
\end{eqnarray*}
Using formulae \eref{F_GV} and \eref{G-append}, we come to
\begin{equation*}
   \left\langle\frac{\partial F_1(V)}{\partial \mu}\right\rangle
    = -T\mathrm{Tr}\, \langle G \rangle + T\mathrm{Tr}\, G_0 = 0.
\end{equation*}
Due to \eref{charge-G0}, we finally obtain
\begin{equation*}
   \left\langle\frac{\partial F_1(V)}{\partial \mu}\right\rangle
    = -T\mathrm{Tr}\, \bar G + N_\mathrm{e} = 0.
\end{equation*}
The latter is the equation for the chemical potential \eref{mu}
of the main text.

\section{Axial symmetry relations}
\label{Appendix-C}
In this section, we prove the axial symmetry of the magnetic susceptibility
in the ferromagnetic state: $\chi^{x}=\chi^{y}$
(with the $z$-axis chosen along the field),
and hence axial symmetry of the fluctuations: $\zeta^x=\zeta^y$.

For given $\mathbf{q}$ and $m$, calculate the susceptibilities
\begin{eqnarray*}
  \chi^{\alpha\beta}_{\mathbf{q}m}
   &\equiv -\frac12\, \frac{\partial^2 F_1^{(2)}}{
    \partial \bigl(\Delta V^{\alpha}_{\mathbf{q}m}\bigr)
    \partial \bigl(\Delta V^{\beta}_{-\mathbf{q}-m}\bigr)}
   \nonumber \\
   &= - \frac{N_\rmd}2 \bigl(1 + 3 \eta \bigr)
    T \sum_{\mathbf{k}n} \sum_{\gamma_1\gamma_2}
    \bar G^{\gamma_1}_{\mathbf{k}n}
    \bar G^{\gamma_2}_{\mathbf{k}-\mathbf{q},\,n-m}
    \Sp \bigl(\sigma^{\gamma_1} \sigma^{\alpha}
    \sigma^{\gamma_2} \sigma^{\beta}\bigr),
\end{eqnarray*}
where $\gamma_1, \gamma_2 = 0,z$. First, using properties of
the Pauli matrices, we find
\begin{equation}\label{chi-spin-matrix}
\chi_{\mathbf{q}m} =
\begin{array}{||ccc||}
   \chi^{xx}_{\mathbf{q}m}& \chi^{xy}_{\mathbf{q}m}& 0 \\
  -\chi^{xy}_{\mathbf{q}m}& \chi^{xx}_{\mathbf{q}m}& 0 \\
   0                      & 0                      & \chi^{zz}_{\mathbf{q}m}
\end{array}
\end{equation}
where
\begin{eqnarray*}
 \chi^{xx}_{\mathbf{q}m}
 &= - N_\rmd \bigl(1 + 3 \eta \bigr) T \sum_{\mathbf{k}n} \Bigl(
  \bar G^{0}_{\mathbf{k}n} \bar G^{0}_{\mathbf{k}-\mathbf{q},\,n-m}
  - \bar G^{z}_{\mathbf{k}n} \bar G^{z}_{\mathbf{k}-\mathbf{q},\,n-m} \Bigr), \\
 \chi^{zz}_{\mathbf{q}m}
 &= - N_\rmd \bigl(1 + 3 \eta \bigr) T \sum_{\mathbf{k}n} \Bigl(
  \bar G^{0}_{\mathbf{k}n} \bar G^{0}_{\mathbf{k}-\mathbf{q},\,n-m}
  + \bar G^{z}_{\mathbf{k}n} \bar G^{z}_{\mathbf{k}-\mathbf{q},\,n-m} \Bigr)
\end{eqnarray*}
and
\begin{equation}\label{chi-xy}
 \chi^{xy}_{\mathbf{q}m}
  = \rmi N_\rmd \bigl(1 + 3 \eta \bigr) T \sum_{\mathbf{k}n} \Bigl(
  \bar G^{0}_{\mathbf{k}n} \bar G^{z}_{\mathbf{k}-\mathbf{q},\,n-m}
  - \bar G^{z}_{\mathbf{k}n} \bar G^{0}_{\mathbf{k}-\mathbf{q},\,n-m} \Bigr).
\end{equation}

Next, we prove that the non-diagonal elements \eref{chi-xy} of matrix
\eref{chi-spin-matrix} are equal to zero. Introducing the new indices,
we rewrite the second part of the sum \eref{chi-xy} as
\begin{equation*}
     \sum_{\mathbf{k}n}
  \bar G^{z}_{\mathbf{k}n} \bar G^{0}_{\mathbf{k}-\mathbf{q},\,n-m}
  =  \sum_{\mathbf{k}'n'} \bar G^{z}_{-\mathbf{k}'+\mathbf{q},\,-n'+m}
  \bar G^{0}_{-\mathbf{k}'-n'}.
\end{equation*}
Using the property $\bar G^{\alpha}_{\mathbf{k}n}=
\bigl( \bar G^{\alpha}_{-\mathbf{k}-n}\bigr)^*$
of the Fourier transformation $\bar G^{\alpha}_{\mathbf{k}n}$ of
the real function $\bar G^{\alpha}_{j}(\tau)$, we have
\begin{equation*}
  \sum_{\mathbf{k}n}
  \bar G^{z}_{\mathbf{k}n} \bar G^{0}_{\mathbf{k}-\mathbf{q},\,n-m}
   = \sum_{\mathbf{k}'n'} \bigl( \bar G^{0}_{\mathbf{k}'n'}
     \bar G^{z}_{\mathbf{k}'-\mathbf{q},\,n'-m} \bigr)^*.
\end{equation*}
Hence for the non-diagonal element \eref{chi-xy} we get
\begin{equation*}
 \chi^{xy}_{\mathbf{q}m}
  = - 2 N_\rmd \bigl(1 + 3 \eta \bigr) T \sum_{\mathbf{k}n} \Im \bigl(
  \bar G^{0}_{\mathbf{k}n} \bar G^{z}_{\mathbf{k}-\mathbf{q},\,n-m} \bigr).
\end{equation*}
Since $\eta$ is real, the latter is also real. On the other hand,
matrix \eref{chi-spin-matrix} is Hermitian, hence $\chi^{xy}_{\mathbf{q}m}$
must be imaginary. Thus $\chi^{xy}_{\mathbf{q}m}=0$, and
the susceptibility $\chi$ is diagonal in the momentum-`frequency'
representation. Formula \eref{sum-qm}
yields the required relation for the fluctuations: $\zeta^x=\zeta^y$.

\section{Derivation of formula \eref{a-coeff-value} for
the correction coefficient $\mbox{\boldmath$\eta$}$}
\label{Appendix-D}

To calculate the correction coefficient \eref{eta-def}:
\begin{equation}\label{eta-def-append}
 \eta
    = \bigl(\Tr(\mathbf{1})\bigr)^{-1}
    \Tr \Bigl(\bar G \bcontraction{}{\Delta V}{\bar G}{\Delta V}
        \Delta V \bar G \Delta V \Bigr),
\end{equation}
where $\Tr(\mathbf{1})=2 N_\mathrm{a} N_\rmd$, we introduce
the Gaussian susceptibility
\begin{equation}\label{chi-prime}
  \grave\chi^{\alpha}_{\mathbf{q}m}
  = - \frac{N_\rmd}2 T \sum_{\mathbf{k}n}
   \sum_{\gamma_1\gamma_2}
   \bar G^{\gamma_1}_{\mathbf{k}n}
   \bar G^{\gamma_2}_{\mathbf{k}-\mathbf{q},\,n-m}
   \Sp \bigl(\sigma^{\gamma_1} \sigma^{\alpha}
   \sigma^{\gamma_2} \sigma^{\alpha}\bigr),
\end{equation}
i.e. the susceptibility $\chi^{\alpha}_{\mathbf{q}m}$ with no account
of high-order terms ($\eta=0$). Then formula \eref{eta-def-append}
 can be rewritten as
\begin{equation}\label{eta-1}
\fl
 \eta
    = - \frac1{N_\mathrm{a} N_\rmd T} \sum_{\mathbf{q}m\alpha}
     \bcontraction{}{\Delta V}{{}^{\alpha}_{\mathbf{q}m}
     \grave\chi^{\alpha}_{\mathbf{q}m}}{\Delta V}
     \Delta V^{\alpha}_{\mathbf{q}m}
     \grave\chi^{\alpha}_{\mathbf{q}m}
     \Delta V^{\alpha}_{\mathbf{-q}-m}
    = - \frac1{N_\mathrm{a} N_\rmd T}
     \sum_{\mathbf{q}m\alpha}
     \grave\chi^{\alpha}_{\mathbf{q}m}
     \langle |\Delta V^{\alpha}_{\mathbf{q}m}|^2 \rangle.
\end{equation}
We approximate \eref{chi-prime} using the single-site Green function
\eref{Green-singlesite} and the quasi-static approximation
(which implies that the main impact to \eref{chi-prime} is due to
the terms with $n-m\approx n$):
\begin{equation}\label{eta-2}
 \grave\chi^{\alpha}_{\mathbf{q}m}
  \approx \grave\chi^{\alpha}_{00}
  = - \frac{N_\rmd}2 T \sum_{n}
  \sum_{\gamma_1\gamma_2}  g^{\gamma_1}_{0n} g^{\gamma_2}_{0n}
  \Sp \left[ \sigma^{\gamma_1} \sigma^{\alpha}
  \sigma^{\gamma_2} \sigma^{\alpha}\right].
\end{equation}
Using \eref{eta-2}, we rewrite \eref{eta-1} as
\begin{equation}\label{eta-3}
  \eta
    \approx - \frac1{N_\rmd N_\mathrm{a} T}
    \sum_{\alpha} \grave\chi^{\alpha}_{00}
    \sum_{\mathbf{q}m}
    \langle |\Delta V^{\alpha}_{\mathbf{q}m}|^2 \rangle
    = - \frac1{N_\rmd T} \sum_{\alpha}
    \grave\chi^{\alpha}_{00} \zeta^{\alpha}.
\end{equation}
Due to the cyclic property of trace and the
anticommutation relations for the Pauli matrices:
$\sigma^{\alpha} \sigma^{\beta}
= (2\delta_{\alpha\beta}-1) \sigma^{\beta} \sigma^{\alpha}$,
we obtain
\begin{equation*}
 \grave\chi^{\alpha}_{00}
  = - N_\rmd T \sum_{n} \bigl( (g^{0}_{0n})^2
  + (2\delta_{\alpha z}-1) (g^{z}_{0n})^2 \bigr).
\end{equation*}
Replacing the summation over `frequencies' by the integration over
the energy variable, we come to
\begin{equation}\label{chi-approx}
 \grave\chi^{\alpha}_{00}
  = - \frac{N_\rmd}{\pi} \int \Im \bigl( (g^{0})^2
  + (2\delta_{\alpha z}-1)(g^{z})^2 \bigr) f\, \rmd\varepsilon.
\end{equation}
Substituting \eref{chi-approx} in \eref{eta-3} and using the axial symmetry
($\zeta^x=\zeta^y$), we get
\begin{eqnarray*}
\fl
   \eta
    = \frac1{\pi T} \biggl( 2\zeta^x\!\!\! \int \Im \bigl( (g^{0})^2
    - (g^{z})^2 \bigr) \, \rmd\varepsilon
    + \zeta^z\!\!\! \int \Im \bigl( (g^{0})^2 + (g^{z})^2 \bigr) \,
    \rmd\varepsilon \biggr).
\end{eqnarray*}
Recalling that $g^0(\varepsilon)=\frac12[g_\uparrow(\varepsilon) +
g_\downarrow(\varepsilon)]$ and
$g^z(\varepsilon)=\frac12[g_\uparrow(\varepsilon) -
g_\downarrow(\varepsilon)]$, we finally obtain
\begin{equation*}
\fl
   \eta \approx \frac1{\pi T} \biggl[\,
   2\zeta^x \!\!\!
   \int \bigl( \Re g_{\uparrow} \Im g_{\downarrow}
   + \Re g_{\downarrow} \Im g_{\uparrow} \bigr) f\, \rmd\varepsilon \,
   + \zeta^z \!\!\!
   \int \bigl( \Re g_{\uparrow} \Im g_{\uparrow}
   + \Re g_{\downarrow} \Im g_{\downarrow} \bigr) f\, \rmd\varepsilon \, \biggr]
\end{equation*}
(see formula \eref{a-coeff-value} of the main text).

\end{document}